\documentclass[doublecol]{epl2}
\usepackage{amsmath}
\usepackage{graphicx}
\usepackage{dcolumn}
\usepackage{bm}
\usepackage{epsfig}
\usepackage[T1]{fontenc}
\usepackage{ae,aecompl}
\usepackage{color}

\begin{document}

\title{Lattice BGK kinetic model for high speed compressible flows: hydrodynamic and nonequilibrium behaviors}
\shorttitle{LBGK model for compressible flows}
\author{Yanbiao Gan$^1$,
\ Aiguo Xu$^2$\footnote{
Corresponding author. E-mail: Xu\_Aiguo@iapcm.ac.cn},
\ Guangcai Zhang$^2$
\ and Yang Yang$^3$
}
\shortauthor{Gan, Xu, Zhang, Yang}

\institute{
$^1$North China Institute of Aerospace Engineering, Langfang 065000,
 P.R.China \\
$^2$National Key Laboratory of Computational Physics, \\
Institute of Applied Physics and Computational Mathematics, P. O.
Box 8009-26, Beijing 100088, P.R.China \\
$^3$China Petroleum Pipeline Bureau, Langfang 065000, P.R.China
}


\abstract{
We present a simple and general approach to formulate the lattice BGK model for high speed compressible flows. The main point consists of two parts: an appropriate discrete equilibrium distribution function (DEDF) $\mathbf{f}^{eq}$ and a discrete velocity model with flexible velocity size.  The DEDF is obtained by $\mathbf{f}^{eq}=\mathbf{C}^{-1}\mathbf{M}$,
where $\mathbf{M}$ is a set of moment of the Maxwellian distribution function,
and $\mathbf{C}$ is the matrix connecting the DEDF and the moments. The numerical components of $\mathbf{C}$ are determined by the discrete velocity model. The calculation of $\mathbf{C}^{-1}$ is based on the analytic solution which is a function of the parameter controlling the sizes of discrete velocity.
The choosing of discrete velocity model has a high flexibility. The specific heat ratio of the system can be flexible.
The approach works for the one-, two- and three-dimensional model constructions.
As an example, we compose a new lattice BGK kinetic model which works not only for recovering the Navier-Stokes equations in the continuum
limit but also for measuring the departure of system from its thermodynamic equilibrium. Via adjusting the sizes of the discrete velocities
the stably simulated Mach number can be significantly increased up to $30$ or even higher.
The model is verified and validated by well-known benchmark tests.
Some macroscopic behaviors of the system due to deviating from thermodynamic equilibrium around
the shock wave interfaces are shown.
}
\maketitle
\section{Introduction}

Compressible flows with high Mach numbers (HMNs) are ubiquitous in many application
fields, ranging from engineering to earth science and even daily life, such
as hydraulic mining, high-pressure water jet cleaning,
dynamics of inertial confinement fusion capsule \cite{ICF}, turbulent dynamics in the Solar convection zone \cite{Solar},
explosive volcanism \cite{EV}, explosion physics and aerospace engineering, etc. However,
due to the complex nature and inherent nonlinearities
of the HMN flows, theoretical solutions and experimental
approaches encounter serious difficulties, and subsequently, the simulation
of compressible flows has become a key tool in both fundamental and applied
research.

The essential characteristic of compressible flows is that, with the increase
of Mach number, the compressibility of flows are more pronounced and the
out-of-equilibrium effects become significant, even dominant. Therefore, how
to exactly describe and capture the nonequilibrium effects is the key issue
of physical modeling. For modeling such a system, as a multiscale approach,
the lattice Boltzmann (LB) method \cite{Succi-Book,Yeomans,Sofonea}, which contains system
information beyond one level of description \cite{Succi-CSE-2001}, has more intrinsic
merits than the traditional hydrodynamic descriptions. The compelling reasons are as below.
Firstly, the LB method is based on the Boltzmann equation which is one of the most
fundamental equations in nonequilibrium statistical physics \cite{Nonideal-Book}.
It naturally inherits the intrinsic characteristics of the latter. Chapman-Enskog
analysis demonstrates that the theoretical framework of LB is self-adaptive
for describing complex systems where the deviations from equilibrium are
spatially and temporally varying. This is a dominating reason why so many
researchers appreciate LB. Secondly, recent studies indicate that the LB
method provides us with an effective way to quantitatively and real-timely study the
departure of the system from its thermodynamic equilibrium \cite{Nonequilibrium},
so it can present helpful information on how to improve the physics modeling from the
macroscopic level. Meanwhile, its superiority is also owed to its linear
convection term, easy implementation of boundary conditions, simple
programming and high efficiency in parallel computing, etc.

In fact, given the great importance of high speed compressible flows in the
aforementioned fields, constructing LB models for this area has been
attempted since the early days of LB research, which yields the
following five approaches, i.e., the adaptive approach by Sun \emph{et al.} \cite{SCH},
the circular function approach by Qu, \emph{et al.} \cite{QuKun}, the
double-distribution-function (DDF) approach by Li, \emph{et al.} \cite{LiQing}, the
additional viscosity single-relaxation-time (AD SRT) approach \cite{Pan2007-Gan2008,Chen2009}
and the multiple-relaxation-time (MRT) approach \cite{Chen-EPL,Chen-PLA} by
our group. In the adaptive approach, particle velocities can vary
with the Mach number and internal energy, and the Maxwellian distribution function (MDF) is replaced by the Kronecker function.
In Qu's approach, a simple circular function that satisfies all needed
statistical relations to recover the Euler equations is introduced to
replace the conventional MDF. In the DDF approach, two distribution functions are utilized:
one for the density and velocity fields, and the other for temperature field. The two distribution functions are coupled with each
other via the equation of state. Those methods successfully recover the target hydrodynamic equations in the continuum limit
and can be used to study the corresponding macroscopic flow behaviors.
In the AD SRT method, the introducing of additional viscosity follows the methodology of traditional computational schemes.
But to investigate the thermodynamic nonequilibrium effects, we have to resort on models where more kinetic processes are reasonably described.
At this side, constructing MRT model is an interesting  methodology.

Due to theoretical simplicity and computational efficiency, how to simulate high speed
compressible flows under the lattice BGK framework without introducing explicit additional dissipation is still a meaningful task.
For such LB models, the construction of appropriate DEDF is of primary importance.
To this end, we suggest to calculate the DEDF via $\mathbf{f}^{eq}=\mathbf{C}^{-1}\mathbf{M}$,
where $\mathbf{M}$ is the set of moment satisfied by the MDF,
and $\mathbf{C}$ is the matrix connecting the DEDF and the moments. The lattice BGK model with such a DEDF
not only recovers the Navier-Stokes equations (NSEs) in the hydrodynamic limit, but also presents deeper insight
into the actual kinetic processes which are beyond the NS description.

\section{Formulation of the discrete equilibrium distribution function}

How to correctly formulate the DEDF $\mathbf{f}^{eq}$ is the foremost essence of the LB method,
because $\mathbf{f}^{eq}$ decides the way and direction the system evolves to.
Here we present a simple and general method to construct DEDF. This approach is based on the
following facts. The kinetic moments that the local DEDF satisfies are bridges and links between
the mesoscopic LB method and the hydrodynamic descriptions at the macroscopic level, which ensure
the correct recovery of the macroscopic equations in the continuum limit. Specifically, Chapman-Enskog
multiscale expansion indicates that, in order to recover the NSEs with flexible
specific-heat ratio from the LB scheme, $\mathbf{f}^{eq}$ should satisfy the following seven kinetic moments \cite{Katato-Watari},
\begin{equation}
\rho =\sum\nolimits_{i=1}^{N}f_{i}^{eq},
\end{equation}
\begin{equation}
\rho u_{\alpha }=\sum\nolimits_{i=1}^{N}f_{i}^{eq}v_{i\alpha }{,}
\end{equation}
\begin{equation}
\rho \left( bT+u_{\alpha }^{2}\right) =\sum\nolimits_{i=1}^{N}f_{i}^{eq}\left(
v_{i\alpha }^{2}+\eta _{i}^{2}\right) ,
\end{equation}
\begin{equation}
P\delta _{\alpha \beta }+\rho u_{\alpha }u_{\beta
}=\sum\nolimits_{i=1}^{N}f_{i}^{eq}v_{i\alpha }v_{i\beta },
\end{equation}
\begin{equation}
\rho \left[ \left( b+2\right) T+u_{\beta }^{2}\right] u_{\alpha
}=\sum\nolimits_{i=1}^{N}f_{i}^{eq}\left( v_{i\beta }^{2}+\eta _{i}^{2}\right)
v_{i\alpha },
\end{equation}
\begin{eqnarray}
\rho \left[ T\left( u_{\alpha }\delta _{\beta \chi }+u_{\beta }\delta
_{\alpha \chi }+u{_{\chi }}\delta _{\alpha \beta }\right) +u_{\alpha
}u_{\beta }u{_{\chi }}\right]   \notag \\
=\sum\nolimits_{i=1}^{N}f_{i}^{eq}v_{i\alpha }v_{i\beta }v_{i\chi },
\end{eqnarray}
\begin{eqnarray}
\rho \left( b+2\right) T{^{2}}\delta _{\alpha \beta }+\left[ \left(
b+4\right) u_{\alpha }u_{\beta }+u_{\chi }^{2}\delta _{\alpha \beta }\right]
\rho T  \notag \\
+\rho u_{\alpha }u_{\beta }u_{\chi }^{2}
=\sum\nolimits_{i=1}^{N}f_{i}^{eq}\left( v_{i\chi }^{2}+\eta _{i}^{2}\right)
v_{i\alpha }v_{i\beta },
\end{eqnarray}
where $\rho$, $T$, $P$($=\rho T$), and $u$ are, respectively, the density, temperature, pressure and velocity. $v_{i}$
is a DVM used to discrete the velocity space. Besides the translational degrees of freedom, $\eta_{i}$ is a free parameter
introducing to describe the $(b-2)$ extra degrees of freedom corresponding to molecular rotation and/or vibration. Here $\eta_{i}=\eta_{0}$
for $i=1$, $\cdots $, $4$, and $\eta_{i}=0$ for $i=5$, $\cdots $, $N$, $N$ is the number of discrete velocities. Subsequently,
the specific-heat ratio can be defined as $\gamma = (b+2)/b$.

To recover the complete NSEs and obtain a higher computational efficiency, we choose the following
two-dimensional DVM which only has sixteen discrete velocities (see fig. 1),
\begin{equation*}
(v_{ix}, v_{iy})=\left\{
\begin{array}{cc}
\mathbf{cyc}:c(\pm 1,0), & \text{for }1\leq i\leq 4, \\
\sqrt{2}c(\pm 1,\pm 1), & \text{for }5\leq i\leq 8, \\
\mathbf{cyc}:2c(\pm 1,0), & \text{ for }9\leq i\leq 12, \\
2\sqrt{2}c(\pm 1,\pm 1), & \text{ }\ \text{for }13\leq i\leq 16,%
\end{array}%
\right.
\end{equation*}
where \textbf{cyc} indicates the cyclic permutation.
The Chapman-Enskog analysis shows that if $\mathbf{f}^{eq}$ satisfies the seven statistical relations in eqs. (1-7),
then the LB equation can recover the NSEs as in ref. \cite{Katato-Watari} by using the above DVM.

%%%%%%%%%%%%%%%%%%%%%%%%%%%%%%%%%%%%%%%%
\begin{figure}[tbp]
{%
\centerline{\epsfig{file=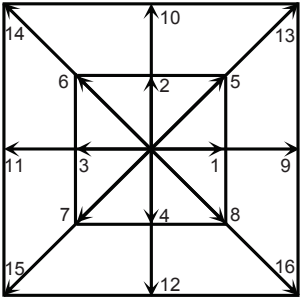,bbllx=0pt,bblly=382pt,bburx=288pt,bbury=672pt,
width=0.22\textwidth,clip=}}}
\caption{ Schematic of the discrete-velocity model.}
\end{figure}
%%%%%%%%%%%%%%%%%%%%%%%%%%%%%%%%%%%%%%%%

Actually, eqs. (1-7) can be written in a matrix form, i.e.,
\begin{equation}
\mathbf{C\times f}^{eq}=\mathbf{M},
\end{equation}
where the bold-face symbols denote
N-dimensional column vectors. Here $\mathbf{f^{eq}}=( f^{eq}_{1},f^{eq}_{2},\cdots
,f^{eq}_{N})^{T}$, $\mathbf{M}=(M_{1}, M_{2},\cdots, M_{N})^{T}$, is the set of moment relations satisfied by the MDF.
$\mathbf{C}=(C_{1},C_{2},\cdots ,C_{N}) ^{T} $, $
C_{i}=(c_{i1}, c_{i2},\cdots, c_{iN})$, and $C_{i}=(1,v_{ix},v_{iy},v_{ix}^{2}+v_{iy}^{2}+\eta
_{i}^{2},v_{ix}v_{iy},v_{ix}^{2},v_{iy}^{2},v_{ix}(v_{ix}^{2}+v_{iy}^{2}+%
\eta _{i}^{2}),v_{iy}(v_{ix}^{2}+v_{iy}^{2}+\eta _{i}^{2}),$
$v_{ix}^{3},v_{iy}^{3},v_{ix}^{2}v_{iy},v_{ix}v_{iy}^{2},v_{ix}v_{iy}(v_{ix}^{2}+v_{iy}^{2}+\eta _{i}^{2}),v_{ix}^{2}(v_{ix}^{2}+v_{iy}^{2}+\eta _{i}^{2}),v_{iy}^{2}(v_{ix}^{2}+v_{iy}^{2}+\eta _{i}^{2}))$. Correspondingly, according to the seven moments, we have
$M_{1}=\rho $, $M_{2}=\rho u_{x}$, $M_{3}=\rho u_{y}$, $M_{4}=\rho (bT+u_{x}^{2}+u_{y}^{2})$, $M_{5}=\rho u_{x}u_{y}$,
$M_{6}=\rho u_{x}^{2}+P$, $M_{7}=\rho u_{y}^{2}+P$, $M_{8}=\rho
u_{x}[(b+2)T+u_{x}^{2}+u_{y}^{2}]$, $M_{9}=\rho u_{y}[(b+2)T+u_{x}^{2}+u_{y}^{2}]$, $M_{10}=\rho
u_{x}(3T+u_{x}^{2})$, $M_{11}=\rho u_{y}(3T+u_{y}^{2})$, $M_{12}=\rho u_{y}(T+u_{x}^{2})$, $M_{13}=\rho
u_{x}(T+u_{y}^{2})$,  $M_{14}=(b+4)\rho Tu_{x}u_{y}+\rho
u_{x}u_{y}(u_{x}^{2}+u_{y}^{2})$, $M_{15}=(b+2)\rho T^{2}+[(b+4)u_{x}^{2}+u_{x}^{2}+u_{y}^{2}]\rho
T+\rho u_{x}^{2}(u_{x}^{2}+u_{y}^{2})$, $M_{16}=(b+2)\rho
T^{2}+[(b+4)u_{y}^{2}+u_{x}^{2}+u_{y}^{2}]\rho T+\rho
u_{y}^{2}(u_{x}^{2}+u_{y}^{2})$.

As a result, $\mathbf{f}^{eq}$ can be calculated from the following way,
\begin{equation}
\mathbf{f}^{eq}={{\mathbf{C}}^{-1}}\mathbf{M},
\end{equation}
where $\mathbf{C}^{-1}$ is the inverse matrix of $\mathbf{C}$, as shown in the Appendix.
Then the system evolves according to the following LB equation:
\begin{equation}
\frac{\partial f_{i}}{\partial t}+v_{i\alpha }\frac{\partial f_{i}}{\partial
r_{\alpha }}=-\frac{1}{\tau }[f_{i}-\mathbf{C}^{-1} \mathbf{M}].
\end{equation}

The approach we presented here has the following advantages over the existing ones.
Compared to the SRT approach \cite{Nonequilibrium,Pan2007-Gan2008,Chen2009,Katato-Watari}
where $\mathbf{f}^{eq}$ is conveniently expanded in terms of macroscopic quantities by only keeping the first relevant orders, this way is more
simple and convenient, since it needs not to solve the complex kinetic moments to give the specific formulations of $\mathbf{f}^{eq}$.
Compared to the MRT approach \cite{Chen-EPL,Chen-PLA}, the physical modeling process is more natural and straightforward,
since the constructions of the transformation matrix and corresponding
distribution function in the kinetic moment space are direct. Furthermore, this scheme has better generalization, since it works for all the one-, two- and three-dimensional model constructions and the choosing of DVM has a high degree of flexibility which results in higher efficiency and better stability of this approach.
In the present model $c$ and $\eta_{0}$ are two free parameters. von Neumann stability
analysis \cite{Pan2007-Gan2008} and numerical experiments demonstrate that these two parameters have great effects
on the stability of the LB model. So they should be varied with the specific problem we studied.
Fortunately, we find that in real simulations if $c$  approaches half of the maximum
velocity of the system $u_{\max}$ and $\eta_{0}$ is equal to or larger than $bc$, the simulations are generally stable.

\section{Numerical Simulations and Analysis}

In this section, several typical benchmarks are carried
out to validate the newly proposed model, including the one- and two-dimensional (1D and 2D) Riemann problems (RPs).
In order to improve the numerical stability, accuracy and efficiency, the third-order implicit-explicit Runge-Kutta finite difference
scheme \cite{LiQing} is utilized to discrete the temporal derivative.
The nonoscillatory and nonfree-parameter dissipation (NND) scheme \cite{NND} is adopted to discrete the spatial derivative
and to capture the discontinuities in compressible flows.

\textit{a) 1D RP: Collision of two strong shocks.}
%%%%%%%%%%%%%%%%%%%%%%%%%%%%%%%%%%%%%%%%
\begin{figure}[tbp]
{%
\centerline{\epsfig{file=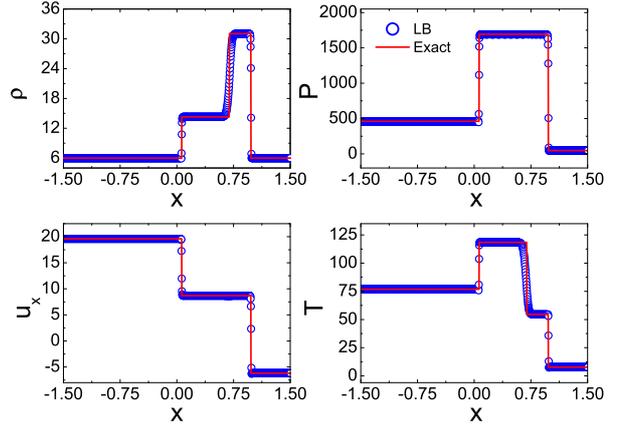,bbllx=11pt,bblly=2pt,bburx=573pt,bbury=408pt,
width=0.44\textwidth,clip=}}}
\caption{Comparisons between LB results and the exact solutions for the collision of two strong shocks, where $t=0.08$ and $\gamma=1.4$.}
\end{figure}
%%%%%%%%%%%%%%%%%%%%%%%%%%%%%%%%%%%%%%%%%

The initial conditions for this test are
\begin{equation*}
\left\{
\begin{array}{c}
(\rho ,T,u_{x},u_{y})|_{L}=(5.99924,76.8254,19.5975,0.0), \\
(\rho ,T,u_{x},u_{y})|_{R}=(5.99242,7.69222,-6.19633,0.0),%
\end{array}%
\right.
\end{equation*}
where subscripts ``L" and ``R" indicate the left and right macroscopic variables of
discontinuity. This is generally regarded as a difficult test. Exact solution contains a
left-shock propagating rightward, a rightward contact discontinuity, and a right-shock that
moves to the right side. Furthermore, the left-shock spreads to the right side very slowly,
which brings additional difficulties to the numerical method. Figure 2 shows comparisons between the
present LB results and the exact solutions for this test, where $t=0.08$ and $\gamma=1.4$.
Parameters used here are $\Delta x=\Delta y= 3 \times 10^{-3}$, $\Delta t=10^{-4}$, $\tau=4\times 10^{-5}$, $c=8.7$ and $\eta_{0}=45$.
The periodic boundary conditions are adopted in the $y$-direction. In the $x$-direction, we set the system at the boundaries
keeps at their corresponding equilibrium states, i.e., $f_{i}=f^{eq}_{i}$. So the macroscopic quantities on the boundary
nodes keep their initial values. The two sets of results have a satisfying agreement. The shock wave and contact
discontinuities are captured well, demonstrating the high accuracy and robustness of the new model. Nevertheless,
we find that the unphysical oscillations at the discontinuities are effectively eliminated and the numerical dissipations
are severely curtailed in the present model than that in the MRT model \cite{Chen-EPL}. This indicates that the problematic
``wall-heating" phenomenon is much weaker. Moreover, the time step $\Delta t$ in the present model is much larger than that
in ref. \cite{Chen-EPL}, so the present model owns a higher computational efficiency for this test.

\textit{b) 1D RP: Colella's explosion wave problem.}
%%%%%%%%%%%%%%%%%%%%%%%%%%%%%%%%%%%%%%%%
\begin{figure}[tbp]
{\centerline{\epsfig{file=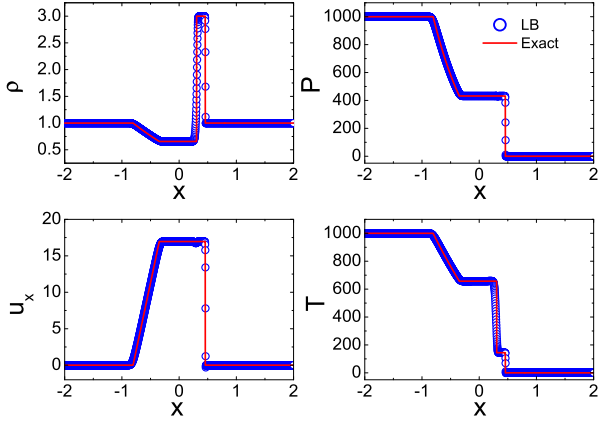,bbllx=11pt,bblly=2pt,bburx=573pt,bbury=403pt,
width=0.44\textwidth,clip=}}}
\caption{Comparisons between LB results and the exact solutions
for Colella's explosion wave test, where $t=0.018$ and $\gamma=2$.}
\end{figure}
%%%%%%%%%%%%%%%%%%%%%%%%%%%%%%%%%%%%%%%%%

For this problem considered, the initial conditions are
\begin{equation*}
\left\{
\begin{array}{c}
(\rho ,T,u_{x},u_{y})|_{L}=(1.0,1000.0,0.0,0.0), \\
(\rho ,T,u_{x},u_{y})|_{R}=(1.0,0.01,0.0,0.0),%
\end{array}%
\right.
\end{equation*}
This is also a difficult and challenging test that commonly used to examine the robustness and precision of numerical methods. The exact solution contains a leftward rarefaction wave, a contact discontinuity and a strong shock. Figure 3 presents comparisons of the present model and theoretical results at $t=0.018$. Here $\Delta x=\Delta y=2 \times 10^{-3}$, $\Delta t=10^{-5}$, $c=20$ and $\eta_{0}=300$. Successful simulation of this test manifests that the present model is applicable to flows with very high ratios of temperature and pressure (up to $10^{5}$).

\textit{c) 1D RP: super-HMN shock problem}
%%%%%%%%%%%%%%%%%%%%%%%%%%%%%%%%%%%%%%%%
\begin{figure}[tbp]
{%
\centerline{\epsfig{file=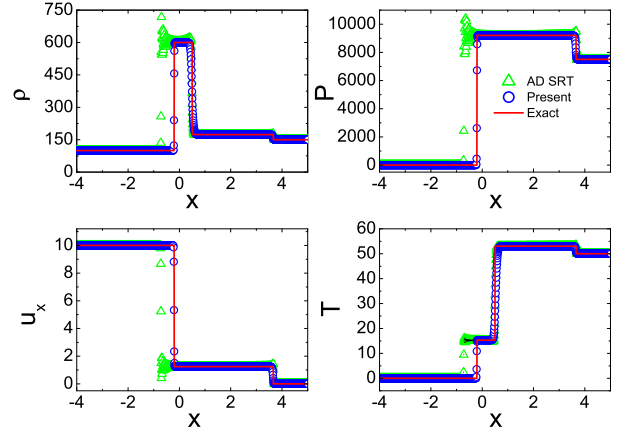,bbllx=11pt,bblly=2pt,bburx=573pt,bbury=408pt,
width=0.44\textwidth,clip=}}}
\caption{Comparisons between the present LB results, the AD SRT results, and the exact solutions for the super-HMN shock problem, where $t=0.4$ and $\gamma=1.4$.}
\end{figure}
%%%%%%%%%%%%%%%%%%%%%%%%%%%%%%%%%%%%%%%%%

To further test the adaptability of the model for HMN problem,
we employe the super-HMN shock tube (The highest Mach number in the system is $Ma=u/\sqrt{\gamma T}=267.26$.) proposed in ref. \cite{Chen2009}.
The initial conditions are described by
\begin{equation*}
\left\{
\begin{array}{c}
(\rho ,T,u_{x},u_{y})|_{L}=(100.0,0.001,10.0,0.0), \\
(\rho ,T,u_{x},u_{y})|_{R}=(150.0,50.0,0.0,0.0),%
\end{array}%
\right.
\end{equation*}
Comparisons between the present model, the AD SRT model in ref. \cite{Chen2009}, and the exact solutions at $t=0.4$ are plotted in fig. 4.
Parameters for this test are $\Delta x=\Delta y=8 \times 10^{-3}$, $\Delta t=10^{-4}$, $\tau=2\times 10^{-5}$, $c=7$ and $\eta_{0}=300$.
Parameters for the AD SRT model are identical to that used in the original publication. It is clear that, simulation results obtained from
the present model agree well with the exact ones, but the AD SRT model cannot accurately capture the position of the shock wave.
Additionally, there exist some spurious numerical oscillations near the sharp discontinuities. This test reminds us that, incorporating
additional physical viscosity can really improve the stability of LB models, but due to the $f^{eq}_{i}$ in ref. \cite{Chen2009} is based
on a low-Mach-number Taylor expansion of MDF, this approach cannot describe very high speed flows because of the insufficient truncation in the
DEDF and the insufficient isotropy in the DVM. However, this limitation is free in our model.

\textit{d) 2D RP: regular reflection and its nonequilibrium characteristics.}
%%%%%%%%%%%%%%%%%%%%%%%%%%%%%%%%%%%%%%%%%%%%%%%%%%%%%%%%%%%%%%%%%%%
\begin{figure}[tbp]
{\centerline{\epsfig{file=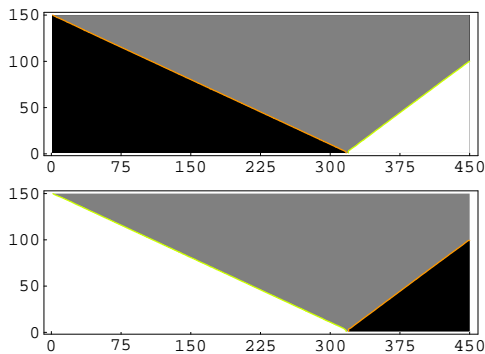,bbllx=119pt,bblly=493pt,bburx=500pt,bbury=770pt,
width=0.36\textwidth,clip=}}}
\caption{ (Color online) Density (upper) and velocity along the $x$-direction (lower) contours of RR on a wall at $t=0.4$. From black to white, the corresponding values increase.}
\label{fig:wide}
\end{figure}
%%%%%%%%%%%%%%%%%%%%%%%%%%%%%%%%%%%%%%%%%%%%%%%%%%%%%%%%%%%%%%%%%%%

It is well known that the reflection of oblique shock waves over a horizontal
plane results in two types of wave configurations, regular reflection (RR) and Mach
reflection (MR). Studies of such shock reflections are of great importance both fundamental
research and engineering applications. For example, there exist strong shock wave reflection
and shock wave interaction phenomenon in hypersonic aircraft, inertial confinement fusion, weapon detonation, etc.
Moreover, it also plays an important role in natural phenomena, such as Supernova explosions.
Up to date, how to describe and analyze the nonequilibrium characteristics of such cases still remains a challenging issue.

Here we simulate a RR process and study its nonequilibrium effects via the present model.
In this test, the incoming shock wave with Mach number $30$ has an angle of $25^{\circ}$ to the wall \cite{Chen2009}.
The computational domain is a rectangle of length 4.5 and height 1.5 divided into a $450 \times 150$ rectangular grids.
Other parameters are $\Delta t=5\times 10^{-5}$, $\tau=2\times 10^{-5}$, $b= 0.858738$, $c=18.0$ and $\eta_{0}=12.0$.
The boundary conditions are composed of a reflecting surface along the bottom boundary, supersonic outflow along the right boundary,
and Dirichlet conditions on the left and upper boundary, respectively,
\begin{equation*}
\left\{
\begin{array}{c}
(\rho ,T,u_{x},u_{y})|_{0,y,t}=(1.0,1/3.329,0.0,0.0), \\
(\rho ,T,u_{x},u_{y})|_{x,1.5,t}=(1.84886,40.0803,27.5399, \\
\text{ \ \ \ \ \ \ \ }-5.27567).
\end{array}%
\right.
\end{equation*}
Figure 5 gives the contours of density and velocity along the $x$-direction at $t=0.4$.
The clear shock reflections on the wall are accordant with the exact solution and results obtained from refs. \cite{Pan2007-Gan2008,Chen-PLA,Chen2009},
so the model has the ability to accurately capture the shock front.

Recently, we define an approach to quantitatively study the
departure of the system from its thermodynamic equilibrium via LB method, i.e, we introduce
%%%%%%%%%%%%%%%%%%%%%%%%%%%%%%%%%%%%%%%%%%%%%%%%%%%%%%%%%%%%%%%%%%%%%%%%%%%%%%%%%%%%%%%%%%%%%%%%%%%%%%%%%%%%%%%%%%%%%%%%%
\begin{equation}
\boldsymbol{\Delta}_m^{*}=\textbf{I}_m^{*}(f_{i})-\textbf{I}_m^{*}(f_{i}^{eq}),~~~~~(m=4,5,6,7)
\label{mom}
\end{equation}
%%%%%%%%%%%%%%%%%%%%%%%%%%%%%%%%%%%%%%%%%%%%%%%%%%%%%%%%%%%%%%%%%%%%%%%%%%%%%%%%%%%%%%%%%%%%%%%%%%%%%%%%%%%%%%%%%%%%%%%%%
as a measure for the deviation of system from its thermodynamic equilibrium,
where the subscript $m$ indicates the $m$-th moment, $\textbf{I}_m^{*}(f_{i})$ and $\textbf{I}_m^{*}(f_{i}^{eq})$ are
central moments calculate by $f_{i}$ and $f_{i}^{eq}$, respectively.
Specifically, $\Delta_{5,\alpha}^*=I_{5,\alpha}^*(f_{i})-I_{5,\alpha}^*(f_{i}^{eq})
=\sum_{i}(f_{i}-f^{eq}_{i})\left[(v_{i\beta}-u_\beta)^{2}+\eta _{i}^{2}\right]
(v_{i\alpha} - u_\alpha)$.
To exhibit the essential advantage of the LB method over traditional ones, we
focus on the nonequilibrium characteristics near the shock wave.

%%%%%%%%%%%%%%%%%%%%%%%%%%%%%%%%%%%%%%%%%%%%%%%%%%%%%%%%%%%%%%%%%%%
\begin{figure}[tbp]
{\centerline{\epsfig{file=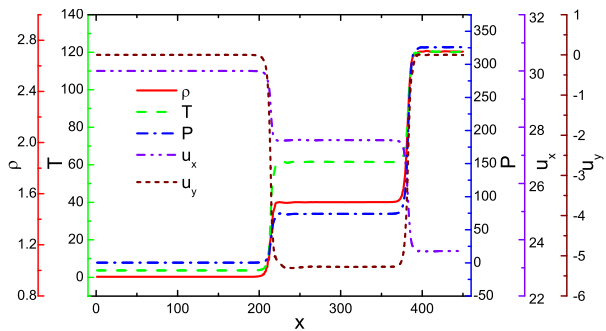,bbllx=6pt,bblly=2pt,bburx=570pt,bbury=307pt,
width=0.44\textwidth,clip=}}}
\caption{ (Color online) Profiles of physical quantities for RR along the horizontal line $y=N_{y}/3$ at $t=0.4$.}
\label{fig:wide}
\end{figure}
%%%%%%%%%%%%%%%%%%%%%%%%%%%%%%%%%%%%%%%%%%%%%%%%%%%%%%%%%%%%%%%%%%%
%%%%%%%%%%%%%%%%%%%%%%%%%%%%%%%%%%%%%%%%%%%%%%%%%%%%%%%%%%%%%%%%%%%
\begin{figure}[tbp]
{\centerline{\epsfig{file=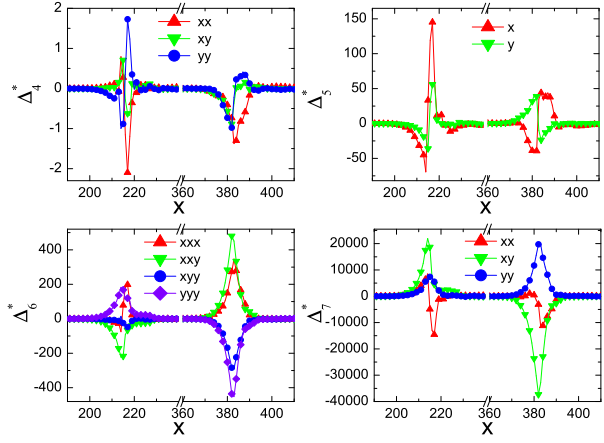,bbllx=6pt,bblly=2pt,bburx=574pt,bbury=419pt,
width=0.44\textwidth,clip=}}}
\caption{ (Color online) Deviations $\boldsymbol{\Delta}_m^{*}$ ($m=4,5,6,7$) for RR along the horizontal line $y=N_{y}/3$ at $t=0.4$.}
\label{fig:wide}
\end{figure}
%%%%%%%%%%%%%%%%%%%%%%%%%%%%%%%%%%%%%%%%%%%%%%%%%%%%%%%%%%%%%%%%%%%

Figure 6 illustrates the profiles of physical quantities for RR along the line $y=N_{y}/3$ at $t=0.4$.
It is found that, there are two shock waves (at about $x=215$ and $381$, respectively) along the horizontal line.
Due to the compression effects of shock waves, density, pressure, and temperature increase sharply. Moreover, as a result of the
strong interactions between the incident shock wave and the reflection shock wave, the second shock is considerably stronger.
Figure 7 presents the corresponding nonequilibrium manifestations, deviations $\boldsymbol{\Delta}_m^{*}$ ($m=4,5,6,7$).
We mention that the nonequilibrium effects are only pronounced around the shock wave interface.
The deviation $\boldsymbol{\Delta}_4^{*}$ associates with the variance of the distribution function.
Its trace associates with the deviation of internal energy calculated from $f_{i}$ and $f^{eq}_{i}$, and
its off-diagonal components associate with the shear effects. From fig. 7, it is interesting to find that,
at the center of the first shock interface (CFSI), the system is nearly in its thermodynamic equilibrium.
At the two sides of the CFSI, the $xx$ and $yy$ components of $\boldsymbol{\Delta}_4^{*}$ deviates from the equilibrium
in opposite ways with the same deviation amplitude, but behave qualitatively similar near the second shock wave zone.
The $xy$ component of $\boldsymbol{\Delta}_4^{*}$ also deviates from zero, demonstrating the shear effects should be
taken into account and the system is not in its dynamic equilibrium. $\boldsymbol{\Delta}_5^{*}$ describes the thermodiffusion.
At both the CFSI and the center of the second shock interface, the $x$ and $y$ components of $\boldsymbol{\Delta}_5^{*}$ nearly approach zero,
and deviate from the equilibrium in opposite directions at each center of the shock zone. The second shock wave is much stronger than the first one,
as a result, the thermodiffusion effects are more obvious near the first shock zone. Different from $\boldsymbol{\Delta}_4^{*}$ and $\boldsymbol{\Delta}_5^{*}$,
 most components of $\boldsymbol{\Delta}_6^{*}$ and $\boldsymbol{\Delta}_7^{*}$ arrive their maxima at the middle of the shock wave interface.

\section{Conclusions and remarks}

In this letter, we have presented a general and simple platform for
constructing LB model for compressible flows with HMN and
arbitrary specific-heat ratio. The key technique of this new scheme is to
inversely calculate $\mathbf{f}^{eq}$ from the kinetic moment relations that
it satisfies. The approach is suitable for constructing LB models in any
dimensions and the choosing of DVM has a high degree of flexibility.
By choosing appropriate parameters controlling the sizes of DVM, the stability of the model can
be significantly enhanced. The new scheme has been validated for a series of one- and two-dimensional numerical benchmarks,
always showing satisfactory agreement with theoretical results and previous results. Some macroscopic behaviors
of the system due to deviating from thermodynamic equilibrium around the shock wave interface are also studied.
Through modifying the BGK collision term according to the way in ref. \cite{CTP-2011},
the model is appropriate for simulating flows with flexible Prandtl number.

We are grateful to the anonymous referees for their valuable comments and suggestions which are very helpful for improving our paper. We
warmly thank Dr. Sauro Succi, Dr. Chuandong Lin for
instructive discussions. This work is partly supported by the Science Foundations of China Academy of
Engineering Physics (under Grant Nos. 2012B0101014 and 2011A0201002),
National Natural Science Foundation of China (under Grant Nos. 11075021,
11071024, and 11202003), Foundation of State Key Laboratory of Explosion
Science and Technology (under Grant No. KFJJ14-1M), Science Foundation of Hebei Province
(under Grant No. A2013409003), and Technology Support Program of LangFang
(under Grant Nos. 2012011021/30, 2011011023).

%%%%%%%%%%%%%%%%%%%%%%%%%%%%%%%%%%%%%%%%%%%%%%%%%%%%%%%%%%%%%%%%%%%%%%%%

\section{Appendix. Specific formulation of the inverse matrix $\mathbf{C}^{-1}$}

$\mathbf{C^{-1}}=(C_{1},C_{2},\cdots, C_{16}) ^{T} $, where $
C_{i}=(c_{i1}, c_{i2},\cdots, c_{i16})$, and

$C_{1}= (0 , 0 , 0 , A , 0 , G , H , N , 0 , -N , 0 , 0 , -N , 0 , T , -T )$;

$C_{2}= (0 , 0 , 0 , A , 0 , H , G , 0 , N , 0 , -N , -N , 0 , 0 , -T , T )$;

$C_{3}= (0 , 0 , 0 , A , 0 , G , H , -N , 0 , N , 0 , 0 , N , 0 , T , -T)$;

$C_{4}= (0 , 0 , 0 , A , 0 , H , G , 0 , -N , 0 , N , N , 0 , 0 , -T , T)$;

$C_{5}= (\frac{2}{3} , \frac{1}{3 c} , \frac{1}{3 c} , B , F , I , I , -\frac{N}{2} , -\frac{N}{2} , P , P , \frac{N}{2} , \frac{N}{2} , -S , \frac{S}{2} , $ $ \frac{S}{2})$;

$C_{6}= (\frac{2}{3} , -\frac{1}{3 c} , \frac{1}{3 c} , B , -F , I , I , \frac{N}{2} , -\frac{N}{2} , -P , P , \frac{N}{2} , -\frac{N}{2} , S , $ $ \frac{S}{2} , \frac{S}{2}$;

$C_{7}= (\frac{2}{3} , -\frac{1}{3 c} , -\frac{1}{3 c} , B , F , I , I , \frac{N}{2} , \frac{N}{2} , -P , -P , -\frac{N}{2} , -\frac{N}{2} , $ $ -S , \frac{S}{2} , \frac{S}{2})$;

$C_{8}= (\frac{2}{3} , \frac{1}{3 c} , -\frac{1}{3 c} , B , -F , I , I , -\frac{N}{2} , \frac{N}{2} , P , -P , -\frac{N}{2} , \frac{N}{2} ,  $ $ S , \frac{S}{2} , \frac{S}{2})$;

$C_{9}= (-\frac{1}{2} , 0 , 0 , D , 0 , J , K , -\frac{N}{8} , 0 , Q , 0 , 0 , R , 0 , U , V)$;

$C_{10}= (-\frac{1}{2} , 0 , 0 , D , 0 , K , J , 0 , -\frac{N}{8} , 0 , Q , R , 0 , 0 , V , U )$;

$C_{11}= (-\frac{1}{2} , 0 , 0 , D , 0 , J , K , \frac{N}{8} , 0 , -Q , 0 , 0 , -R , 0 , U , V)$;

$C_{12}= (-\frac{1}{2} , 0 , 0 , D , 0 , K , J , 0 , \frac{N}{8} , 0 , -Q , -R , 0 , 0 , V , U)$;

$C_{13}= (\frac{1}{12} , -\frac{1}{24 c} , -\frac{1}{24 c} , E , -\frac{F}{16} , L , L , \frac{N}{16} , \frac{N}{16} , -\frac{P}{8} , -\frac{P}{8} , $ $ -\frac{R}{2} , -\frac{R}{2} , \frac{S}{4} , \frac{S}{4} , \frac{S}{4})$;

$C_{14}= (\frac{1}{12} , \frac{1}{24 c} , -\frac{1}{24 c} , E , \frac{F}{16} , L , L , -\frac{N}{16} , \frac{N}{16} , \frac{P}{8} , -\frac{P}{8} , -\frac{R}{2} , $ $ \frac{R}{2} , -\frac{S}{4} , \frac{S}{4} , \frac{S}{4})$;

$C_{15}= (\frac{1}{12} , \frac{1}{24 c} , \frac{1}{24 c} , E , -\frac{F}{16} , L , L , -\frac{N}{16} , -\frac{N}{16} , \frac{P}{8} , \frac{P}{8} , \frac{R}{2} , \frac{R}{2} , $ $ \frac{S}{4} , \frac{S}{4} , \frac{S}{4})$;

$C_{16}= (\frac{1}{12} , -\frac{1}{24 c} , \frac{1}{24 c} , E , \frac{F}{16} , L , L , \frac{N}{16} , -\frac{N}{16} , -\frac{P}{8} , \frac{P}{8} , \frac{R}{2} , $ $ -\frac{R}{2} , -\frac{S}{4} , \frac{S}{4} , \frac{S}{4})$,\\
with
$A=\frac{1}{4\eta _{0}^{2}}$, $B=\frac{-21c^{2}-\eta _{0}^{2}}{48c^{2}\eta
_{0}^{2}}$, $D=\frac{7c^{2}+\eta _{0}^{2}}{32c^{2}\eta _{0}^{2}}$,
$E=\frac{-3c^{2}-\eta _{0}^{2}}{96c^{2}\eta _{0}^{2}}$, $F=\frac{1}{3c^{2}}$, $G=
\frac{3c^{2}-5\eta _{0}^{2}}{4\eta _{0}^{2}\left( \eta
_{0}^{2}-3c^{2}\right)}$, $H=\frac{3\left( c^{2}+\eta _{0}^{2}\right)}{4\eta _{0}^{2}\left( \eta
_{0}^{2}-3c^{2}\right)}$, $I=\frac{21c^{2}-11\eta _{0}^{2}}{48c^{2}\eta
_{0}^{2}}$, $J=\frac{21c^{4}-32c^{2}\eta _{0}^{2}+11\eta _{0}^{4}}{32c^{2}\eta
_{0}^{2}\left( \eta _{0}^{2}-3c^{2}\right)}$, $K=\frac{21c^{4}-36c^{2}\eta
_{0}^{2}+7\eta _{0}^{4}}{32c^{2}\eta _{0}^{2}\left( \eta
_{0}^{2}-3c^{2}\right)}$, $L=\frac{3c^{2}-5\eta _{0}^{2}}{96c^{2}\eta _{0}^{2}}$, $N=\frac{1}{2c\eta
_{0}^{2}}$, $P=\frac{3c^{2}-\eta _{0}^{2}}{12c^{3}\eta _{0}^{2}}$,
$Q=\frac{c^{2}+\eta _{0}^{2}}{16c^{3}\eta _{0}^{2}}$, $R=\frac{c^{2}-\eta
_{0}^{2}}{16c^{3}\eta _{0}^{2}}$, $S=\frac{1}{24c^{4}}$, $T=\frac{1}{4c^{2}\left( \eta _{0}^{2}-3c^{2}\right)}$,
$U=\frac{c^{2}-\eta _{0}^{2}}{32c^{4}\left( \eta _{0}^{2}-3c^{2}\right)}$,
$V=\frac{5c^{2}-\eta _{0}^{2}}{32c^{4}\left( \eta _{0}^{2}-3c^{2}\right)}$.

\end{document}